\documentclass{emulateapj}

\usepackage{graphicx} 

\newcommand{\numax}{\mbox{$\nu_{\rm max}$}}
\newcommand{\Dnu}{\mbox{$\Delta \nu$}}

\newcommand{\muHz}{\mbox{$\mu$Hz}}
\newcommand{\kep}{\mbox{\textit{Kepler}}}

\slugcomment{accepted for publication in ApJ}

\shorttitle{Testing Asteroseismic Scaling Relations}
\shortauthors{D. Huber et al.}

\begin{document}

\title{Testing Scaling Relations for Solar-Like Oscillations from the Main Sequence to Red Giants using 
\textit{Kepler} Data}

\author{
D.~Huber\altaffilmark{1}, 
T.~R.~Bedding\altaffilmark{1}, 
D.~Stello\altaffilmark{1}, 
S.~Hekker\altaffilmark{2,3}, 
S.~Mathur\altaffilmark{4}, 
B.~Mosser\altaffilmark{5}, 
G.~A.~Verner\altaffilmark{3,6}, 
A.~Bonanno\altaffilmark{7}, 
D.~L.~Buzasi\altaffilmark{8}, 
T.~L.~Campante\altaffilmark{9,10}, 
Y.~P.~Elsworth\altaffilmark{3}, 
S.~J.~Hale\altaffilmark{3}, 
T.~Kallinger\altaffilmark{11,12}, 
V.~Silva Aguirre\altaffilmark{13}, 
W.~J.~Chaplin\altaffilmark{3}, 
J.~De~Ridder\altaffilmark{14}, 
R.~A.~Garc\'\i a\altaffilmark{15}, 
T.~Appourchaux\altaffilmark{16}, 
S.~Frandsen\altaffilmark{10}, 
G.~Houdek\altaffilmark{12}, 
J.~Molenda-\.Zakowicz\altaffilmark{17}, 
M.~J.~P.~F. G. Monteiro\altaffilmark{9}, 
J.~Christensen-Dalsgaard\altaffilmark{10}, 
R.~L.~Gilliland\altaffilmark{18}, 
S.~D.~Kawaler\altaffilmark{19}, 
H.~Kjeldsen\altaffilmark{10}, 
A.~M.~Broomhall\altaffilmark{3}, 
E.~Corsaro\altaffilmark{7}, 
D.~Salabert\altaffilmark{20}, 
D.~T.~Sanderfer\altaffilmark{21}, 
S.~E.~Seader\altaffilmark{22}, and 
J.~C.~Smith\altaffilmark{22} 
}
\altaffiltext{1}{Sydney Institute for Astronomy (SIfA), School of Physics, University of Sydney, NSW 2006, Australia; \mbox{dhuber@physics.usyd.edu.au}}
\altaffiltext{2}{Astronomical Institute 'Anton Pannekoek', University of Amsterdam, Science Park 904, 1098 XH Amsterdam, The Netherlands}
\altaffiltext{3}{School of Physics and Astronomy, University of Birmingham, Birmingham B15 2TT, UK}
\altaffiltext{4}{High Altitude Observatory, NCAR, P.O. Box 3000, Boulder, CO 80307, USA}
\altaffiltext{5}{LESIA, CNRS, Universit\'e Pierre et Marie Curie, Universit\'e Denis, Diderot, Observatoire de Paris, 92195 Meudon cedex, France}
\altaffiltext{6}{Astronomy Unit, Queen Mary University of London, Mile End Road, London E1 4NS, UK}
\altaffiltext{7}{INAF Osservatorio Astrofisico di Catania, Italy}
\altaffiltext{8}{Eureka Scientific, 2452 Delmer Street Suite 100, Oakland, CA 94602-3017, USA}
\altaffiltext{9}{9 Centro de Astrof\'{\i}sica and Faculdade de Ci\^encias, Universidade do Porto, Rua das Estrelas, 4150-762 Porto, Portugal}
\altaffiltext{10}{Danish AsteroSeismology Centre (DASC), Department of Physics and Astronomy, Aarhus University, DK-8000 Aarhus C, Denmark}
\altaffiltext{11}{Department of Physics and Astronomy, University of British Columbia, Vancouver, Canada}
\altaffiltext{12}{Institute of Astronomy, University of Vienna, 1180 Vienna, Austria}
\altaffiltext{13}{Max-Planck-Institut f\"ur Astrophysik, Karl-Schwarzschild-Str.~1, 85748 Garching, Germany}
\altaffiltext{14}{Instituut voor Sterrenkunde, K.U.Leuven, Belgium}
\altaffiltext{15}{Laboratoire AIM, CEA/DSM-CNRS, Universit\'e Paris 7 Diderot, IRFU/SAp, Centre de Saclay, 91191, Gif-sur-Yvette, France}
\altaffiltext{16}{Institut d'Astrophysique Spatiale, UMR 8617, Universite Paris Sud, 91405 Orsay Cedex, France}
\altaffiltext{17}{Astronomical Institute of the University of Wroc{\l}aw, ul. Kopernika 11, 51-622 Wroc{\l}aw, Poland}
\altaffiltext{18}{Space Telescope Science Institute, 3700 San Martin Drive, Baltimore, Maryland 21218, USA}
\altaffiltext{19}{Department of Physics and Astronomy, Iowa State University, Ames, IA 50011 USA}
\altaffiltext{20}{Universit\'e de Nice Sophia-Antipolis, CNRS, Observatoire de la C\^ote d'Azur, BP 4229, 06304 Nice Cedex 4, France}
\altaffiltext{21}{NASA Ames Research Center, MS 244-30, Moffett Field, CA 94035, USA}
\altaffiltext{22}{SETI Institute/NASA Ames Research Center, MS 244-30, Moffett Field, CA 94035, USA}

\begin{abstract}
We have analyzed solar-like oscillations in $\sim$1700 stars observed by the 
\textit{Kepler} Mission, spanning from the main-sequence to the red clump. 
Using evolutionary models, we test asteroseismic scaling relations for the frequency of maximum 
power ($\nu_{\rm max}$), the large frequency separation ($\Delta\nu$) and oscillation 
amplitudes. We show that the difference of the $\Delta\nu$-$\nu_{\rm max}$ relation for unevolved and 
evolved stars can be explained by different distributions in effective temperature and stellar 
mass, in agreement with what is expected from scaling relations. For 
oscillation amplitudes, we show that neither $(L/M)^s$ scaling nor the revised scaling relation by 
Kjeldsen \& Bedding (2011) is accurate for red-giant stars, and demonstrate that a revised scaling relation 
with a separate luminosity-mass dependence can be used to calculate amplitudes from the 
main-sequence to red-giants to a precision of $\sim$25\%. The residuals show an offset particularly 
for unevolved stars, suggesting that an additional physical dependency is necessary to fully reproduce 
the observed amplitudes.
We investigate correlations between amplitudes and stellar activity, and find evidence
that the effect of amplitude suppression is most pronounced for subgiant stars. 
Finally, we test the location of the cool edge of the instability strip in the 
Hertzsprung-Russell diagram using solar-like oscillations and find the 
detections in the hottest stars compatible with 
a domain of hybrid stochastically excited and opacity driven pulsation.
\end{abstract}


\section{Introduction}

Empirical relations connecting observable quantities with physical parameters of stars 
are of fundamental importance for many fields of stellar astrophysics, with classical examples 
including the colour-temperature scale for late-type stars \citep[see, e.g.,][]{flower,casagrande} 
and the period-luminosity relation for Cepheid variables \citep{leavitt}. The calibration 
of these relations relies either on the direct measurement of stellar properties 
(e.g. through trigonometric parallaxes and interferometry) or the comparison of observations with 
stellar models. In both cases measurements spanning an extended range in parameter space (such as 
temperature, luminosity and metallicity) are required.

One of the most powerful methods to constrain fundamental 
properties of field stars is asteroseismology \citep[see, e.g.,][]{B+G94, ChD2004, thebook}. The 
connection of 
global asteroseismic observables to fundamental stellar properties was calibrated by 
\citet[][hereafter KB95]{KB95} by introducing relations that scale stellar properties from the observed 
values of the Sun. The scaling relations have been used extensively both for the direct inference of 
stellar properties 
\citep[e.g.,][]{stello_radius,kallinger_rg,mosser10,kallinger10,chaplin_science,hekker_public,hekker_cluster,silva} 
as well as for calculating pulsation amplitudes of main-sequence and subgiant stars
\citep[e.g.,][]{michel_science,bonanno,mathur_corot,huber_procyon,chaplin_apj2} and red-giant stars 
\citep{edmonds96,gilliland08,stello09b,mosser10,HBS10,baudin,stello_cluster}. The three 
global observables discussed in this paper are the frequency of maximum power 
(\numax), the large frequency separation (\Dnu) and the mean oscillation amplitude per 
radial mode.

Tests of the scaling relations for these observables over a wide range of parameter space has so far 
been hampered by the relative sparsity of detections of solar-like oscillations.
The success of the \kep\ mission has changed this picture, with the detection of oscillations 
in thousands of stars covering a large part of the low-mass region in the H-R diagram 
\citep{gilliland10}. In this paper, we aim to use the large number of detections by 
\kep\ to test scaling relations for stars ranging from the main-sequence to He-core burning red-giants.

\section{Data Analysis and Models}

The \kep\ space telescope was launched in March 2009 with the primary goal of finding Earth-like
planets orbiting solar-like stars through the detection of photometric transits. \kep\ monitors the 
brightness of stars at two sampling rates, either in 29.4\,min 
(long cadence) or in 58.8\,sec (short cadence) intervals \citep{sc,lc}. For asteroseismic studies 
of solar-like oscillators, 
the former are primarily used to study
oscillations in red-giant stars, and short cadence data are used to measure the more rapid 
oscillations in main-sequence and subgiant stars. While red giants observed in long cadence have 
been continuously monitored since the launch of the mission, the number of short cadence slots is 
restricted due to bandwidth limitations. Short cadence slots have therefore so far been primarily 
used to survey a large number of stars for a 
period of one month each. 

\kep\ observations are subdivided into quarters, starting with 
the initial commissioning run (10\,d, Q0), followed by a short first quarter (34\,d, Q1) and 
subsequent full quarters of 90\,d length. Our studies are based on \kep\ data spanning from Q0 to Q6 
for long cadence data and Q0 to Q4 for short cadence data. 
Our final sample contains 1686 stars, of which 1144 have been observed at long cadence, 
mostly since the launch of the mission ($\sim\,500$ days), and 542 at short cadence for a typical length of one
month. We have used \kep\ raw data, which were reduced in the manner described by \citet{garcia} and 
analyzed using several automated analysis methods 
\citep{bonanno,campante,hekker09b,HSB09,kallinger_rg,karoff,mathur10,mosser09,mosser_universal,verner11}. 
We refer the reader to \citet{hekker_comp} and \citet{verner} for an extensive comparison of the 
results provided by these methods.
Unless otherwise mentioned, all results presented here are based on the method by \citet{HSB09}. 
We have only retained results for stars with at least one matching pipeline result 
within 10\% and 5\% of the determined \numax\ and \Dnu\ value, respectively. The same outlier 
rejection procedure was repeated for all methods that returned results for both long cadence 
and short cadence data, and these datasets have then been used to validate all results and conclusions 
reported in this paper. Note that all amplitudes shown in this paper have been normalized to recover 
the full sine-amplitude of an injected signal (commonly referred to as peak-scaling). 
All amplitudes shown have been calculated using the method described by \citet{KB08} with 
$c=3.04$ \citep{bedding_procyon} to convert to amplitude per radial mode.

Uncertainties on \numax, \Dnu\ and amplitudes reported in this paper were 
estimated using Monte-Carlo simulations by generating
synthetic power spectra
following a $\chi^{2}$ distribution with two degrees of freedom 
with expected values corresponding to the observed power density levels.  
For each star, the method by \citet{HSB09} was  repeated on each synthetic dataset and the standard 
deviation of the distribution after 500 iterations was taken as an estimate of the 
uncertainty. The typical relative uncertainties obtained using this method for 
long cadence and short cadence data are 3\% and 4\% for \numax, 1\% and 3\% for \Dnu, and 
7\% and 11\% for the amplitude. These estimates agree with the results obtained by
\citet{hekker_comp} and \citet{verner}. For our analysis, only stars with uncertainties lower 
than 20\% in \numax, 10\% in \Dnu\ and 50\% in amplitude were retained. 
Note that oscillation amplitudes for all stars observed in long-cadence with 
$\numax>200\muHz$ have been removed from our analysis due to the difficulty of estimating the 
noise level close to the Nyquist frequency of 283\muHz.

The asteroseismic relations discussed in this paper rely on scaling from 
observed values of the Sun. To ensure that these reference values 
are consistent with our analysis method, we used 111 30-day subsets of data collected by the VIRGO 
instrument \citep{frohlich} aboard the SOHO spacecraft spanning from 1996 to 2005 and
analyzed them in the same way as the \kep\ data. This yielded solar reference values of 
$\nu_{\rm max,\sun} = 3090\pm30 \muHz$ and $\Delta\nu_{\sun}=135.1\pm0.1 \muHz$, which are consistent 
with previously quoted values in the literature. We determined the oscillation amplitude to be 
$A_{\sun}=4.4\pm0.3$\,ppm for the VIRGO green channel ($\lambda=500$\,nm) which,  
using the approximation by KB95, translates into a solar bolometric amplitude of 
$A_{\sun,\rm bol}=3.5\pm0.2$\,ppm. This is in excellent agreement with the solar reference value of 
3.6\,ppm established by \citet{michel_ref}, and we have adopted 
$A_{\sun,\rm bol}=3.6$\,ppm for the remainder of this paper.

In the absence of directly determined fundamental properties for most of the \kep\ stars, 
a test of asteroseismic scaling relations relies on the comparison of 
observations with models. In our study, we use canonical BaSTI evolutionary 
models \citep{basti} with a solar-scaled distribution of heavy elements \citep{GN93}. Mass loss in 
BaSTI models is characterized according to the Reimers law \citep{reimers}, and we used 
models with the mass loss parameter set to the commonly used value $\eta=0.4$ 
\citep[see, e.g.,][]{fusipecci}. Note that mass loss for these models is only significant 
in the red giant phase of stellar evolution.

\section{Scaling relations for \numax\ and \Dnu}

\citet{brown91} first argued that the frequency of maximum power (\numax) for Sun-like 
stars should scale with the acoustic cut-off frequency. KB95 used this assumption to relate 
\numax\ to stellar properties as follows:

\begin{equation}
\nu_{\rm max} \approx \frac{ M/M_{\sun}(T_{\rm eff}/T_{\rm eff,\sun})^{3.5}}{L/L_{\sun}} \nu_{\rm max,\sun} \: .
\label{equ:nmax}
\end{equation}

\noindent
This scaling relation has since been found to work well both observationally
\citep[see, e.g.,][]{kb03_review,stello08,bedding_canary} as well as theoretically \citep[see, e.g.,][]{chaplin08,belkacem}.

The mean large frequency separation (\Dnu) between modes of consecutive radial overtone and equal
spherical degree 
is directly related to the sound travel time across the stellar diameter, and is therefore sensitive 
to the mean stellar density \citep{ulrich86}. This is expressed in the following scaling relation:

\begin{equation}
\Delta\nu \approx \frac{(M/M_{\sun})^{0.5}(T_{\rm eff}/T_{\rm eff,\sun})^{3}}{(L/L_{\sun})^{0.75}}
\Delta\nu_{\sun} \: .
\label{equ:dnu}
\end{equation}

 \noindent
Note that in our analysis \Dnu\ is measured as the mean spacing of all detectable modes around the 
value of \numax\ in the power spectrum.

\begin{figure}
\begin{center}
\resizebox{\hsize}{!}{\includegraphics{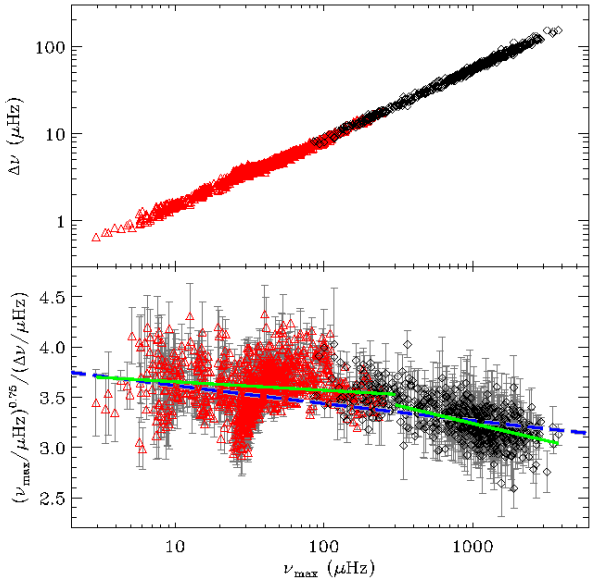}}
\caption{Upper panel: \Dnu\ versus \numax\ for the entire sample of \kep\ stars. Red triangles show
stars observed in long cadence, while black diamonds are stars observed in short cadence. Lower panel: Same as
upper panel, but with the luminosity dependence removed by raising \numax\ to the power of 0.75. Green lines
show 
power law fits to the \Dnu-\numax\ relation for two different intervals of \numax\ (see text). 
The blue dashed line shows the relation by \citet{stello09}.}
\label{fig:02}
\end{center}
\end{figure}

It has been well established for both main-sequence and red-giant stars that \numax\ and \Dnu\ follow 
a power law relation \citep{stello09,hekker09,mosser10,hekker_public,hekker_cluster}:
 
\begin{equation}
\Delta\nu = \alpha (\nu_{\rm max}/\mu\rm Hz)^\beta \: .
\label{equ:dnunumax}
\end{equation}

\noindent
Figure 1 shows this relation for the entire Kepler sample in our analysis. 
Although the relation appears to be constant over several orders of magnitude, \citet{mosser10} and 
\citet{HBS10} noted that 
the slope is different for red-giant and main-sequence stars. 
This can be illustrated more clearly by removing the luminosity dependence by raising 
\numax\ to the power of 0.75, yielding

\begin{equation}
\frac{(\numax/\muHz)^{0.75}}{\Dnu/\muHz} \propto \left(\frac{M}{M_{\sun}}\right)^{0.25} \left(\frac{T_{\rm
eff}}{T_{\rm eff,\sun}}\right)^{-0.375} \: .
\label{equ:dnunumax2}
\end{equation}
 
 \begin{table}
\begin{center}
\caption{Coefficients of the \Dnu-\numax\ relation.}
\begin{tabular}{l c c c c}        
\hline         
\hline
Method			&	$ \alpha,\beta$	& 		$\alpha,\beta$		& \# of stars	\\ 
				&	$\numax<300\muHz$	&	$\numax>300\muHz$		& 		\\
\hline
A2Z		& 	0.259(3),0.765(2)		&	0.25(1),0.779(7)	&	919,257		\\			
COR		& 	0.267(3),0.761(2)		&	0.23(1),0.789(6) 	&	1150,415		\\
OCT		& 	0.263(3),0.763(2)		&	0.20(1),0.811(7) 	&	1082,281		\\
SYD		& 	0.267(2),0.760(2)		&	0.22(1),0.797(5)	&	1228,458		\\
\hline
\end{tabular} 
\end{center}
A2Z - \citet{mathur10}, 
COR - \citet{mosser09,mosser_universal}, OCT - \citet{hekker09b}, 
SYD - \citet{HSB09}. \newline
\label{tab:dnunumax} 
\end{table}

The lower panel of Figure 1 displays this ratio as function of \numax. The distribution shows a 
prominent diagonal structure around $\numax \sim 20-50\muHz$, which we identify as the 
red clump, comprising He-core 
burning red giant stars \citep[see][]{bedding_nature,mosser_mixed,mosser_energy}.
It is evident that the power law becomes steeper as \numax\ increases. To measure this effect, 
we fitted Equation (\ref{equ:dnunumax}) to the sample in two groups subdivided at $\numax=300\muHz$, 
which roughly marks the transition from low-luminosity red giants to subgiants. 
The best-fitting power laws for the SYD pipeline are shown as solid green lines in the 
lower panel of Figure \ref{fig:02}, and the results for each pipeline with values for both long- and 
short cadence data are listed in Table 1. The difference in the 
coefficients between red giant and main-sequence stars 
is significant, and should be noted when using the relation for 
determining \Dnu\ from \numax. For \numax\ close to the solar value, for example, the use of a 
power-law relation calibrated to red-giant stars would lead to an underestimation 
in \Dnu\ by $\sim$ 10\%. For comparison, the blue dashed line in the lower panel of Figure \ref{fig:02} 
shows the relation derived 
by \citet{stello09} using a sample including both main-sequence and red-giant stars.

How do our observed values of \numax\ and \Dnu\ compare with evolutionary models? 
The upper panel of Figure 2 
compares the observed 
distributions with solar-metallicity models with masses 1.0$M_{\sun}$, 1.3$M_{\sun}$
and 2.0$M_{\sun}$ (solid lines), which roughly correspond to the lower bound, median 
and upper bound of the mass 
distribution derived by \citet{kallinger_rg} and \citet{chaplin_science}.
The model values for \numax\ and \Dnu\ have been calculated using Equations 
(\ref{equ:nmax}) and (\ref{equ:dnu}). 
Additionally, we have color-coded the mass of each star calculated using 
Equations (\ref{equ:nmax}) and (\ref{equ:dnu}) with the observed values of \numax\ and \Dnu\, 
adopting the effective
temperature listed in the \kep\ Input Catalog \citep[KIC,][]{kic}\footnote[1]{Note that for 39 stars 
in our list, no effective temperatures were available 
in the KIC, and we omitted those stars for the remainder of our analysis.}. The 
observed and theoretical masses in this plot essentially correspond to a  
comparison of the so-called 
direct method and grid-based method of estimating asteroseismic masses \citep[see, e.g.,][]{gai}, 
assuming solar metallicity.

\begin{figure}
\begin{center}
\resizebox{\hsize}{!}{\includegraphics{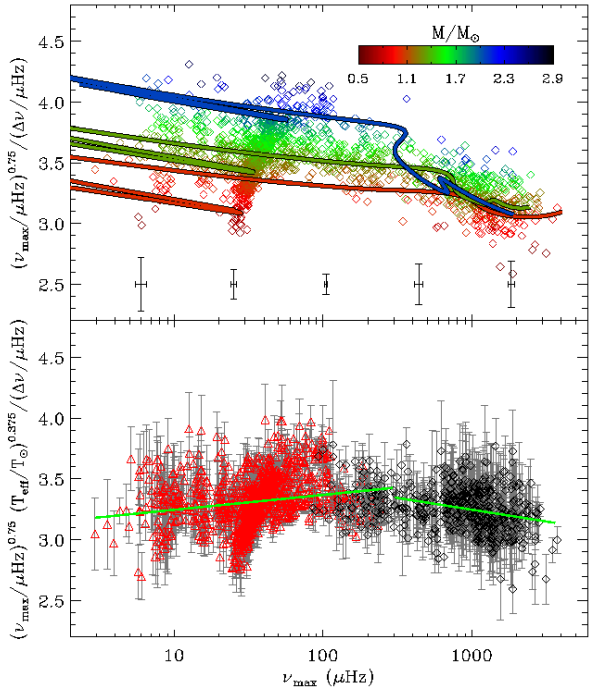}}
\caption{Upper panel: Observed values of $\nu_{\rm max}^{0.75}$/\Dnu\ versus \numax\ (symbols)
compared to solar-metallicity ($Y=0.273$, $Z=0.0198$) models with 1$M_{\sun}$, 
1.3$M_{\sun}$ and 2.0$M_{\sun}$ (solid lines). 
Asteroseismic masses are color-coded as indicated in the plot.
Typical error bars for different ranges of \numax\ are indicated near the bottom of the plot.
Lower panel: Same as top panel but with the effective temperature dependence removed (see Equation 
(\ref{equ:dnunumax2})) and omitting model tracks. Symbol types and colors are the same as in Figure \ref{fig:02}.
The green solid lines show linear fits to the linear-log plot for the same intervals of \numax\ as 
in Figure \ref{fig:02}.}
\label{fig:03}
\end{center}
\end{figure}

As noted by \citet{kallinger10} and 
\citet{HBS10}, the spread in stellar mass is pronounced on the red giant branch, while for less-evolved stars 
the spread in observations is weaker and the models almost overlap. 
The overall agreement between the models and the data is very good, and we do not observe any 
significant offset of the models with respect to the observations. 
It is also remarkable how well the models of 
different masses track the He-core-burning red clump. 
Although the  spread of data points about the models for main-sequence 
stars can be explained by measurement uncertainties, we note that stellar population models presented by 
\citet{silva} yield 
evidence that the \kep\ sample is on average metal-poor, which can have a some 
impact on the \Dnu-\numax\ relation of main-sequence stars \citep[see][]{HBS10}. 
As noted by \citet{silva}, however, the 
measurement uncertainties are currently too large to test a metallicity influence on the scaling 
relations.

What causes the different \Dnu-\numax\ power-law relations for evolved and unevolved stars?
The scaling relation in Equation (\ref{equ:dnunumax2}) suggests that the change in slope for the 
unevolved stars must 
be partially due to a variation in effective temperature. To test this hypothesis, we 
again used the effective
temperatures listed in the \kep\ Input Catalog to model this dependency using Equation
(\ref{equ:dnunumax2}). The result is shown in the lower panel of 
Figure 2. As expected, correcting for the higher average effective temperatures of 
main-sequence stars compared to red giants removes the gradient in the distribution. 
Subdividing the sample again at 
$\numax = 300\muHz$, we find a positive slope for the red giant sample and a negative slope for 
main-sequence stars (solid green lines). These variations are qualitatively in agreement with  
different mass distributions in the sample: while for red giants \numax\ is correlated to 
stellar mass for He-core burning stars \citep{mosser_mixed,mosser_energy}, main-sequence stars with 
higher \numax\ are generally low-mass stars \citep[see, e.g.,][]{chaplin_science}.
Without mass estimates independent of asteroseismic scaling relations, however, it is not 
possible to make further quantitative conclusions about the distribution.

\section{Amplitudes}

\subsection{The L/M Scaling Relation}

KB95 suggested that model predictions by \citet{CDF83} implied a 
scaling for velocity amplitudes of 

\begin{equation}
A_{\rm vel} \propto \left(\frac{L}{M} \right)^{s} \: ,
\label{equ:amps}
\end{equation}

\noindent
with $s=1$. They further argued that the oscillation amplitude $A_{\lambda}$ observed in photometry 
at a wavelength $\lambda$ is related to the velocity amplitude:

\begin{equation}
A_{\lambda} \propto \frac{v_{\rm osc}}{\lambda T_{\rm eff}^{r}} \: .
\label{equ:amps2}
\end{equation}

\noindent
The exponent $s$ has since been revised 
in the range of roughly $s=0.7-1.3$ both 
theoretically \citep{houdek99,houdek06,SGT2007}, as well as observationally using red-giant stars 
\citep{gilliland08,dziem10,mosser10,stello10}, main-sequence stars \citep{verner} and an ensemble of main-sequence and 
red-giant stars \citep{baudin}. 
The value for $r$ has so far 
been chosen to be either $r=1.5$ (assuming adiabatic oscillations) or $r=2.0$ (the best-fitting 
coefficient found by KB95 for classical pulsators).

Since our sample spans a large range in effective temperature, we must account for the spectral 
response of the \kep\ bandpass to compare predictions made with Equation (\ref{equ:amps2}) 
to our observations. To do so, we used the expression for the bolometric amplitude given by 
KB95 and converted these values to amplitudes observed in the \kep\ bandpass, 
as follows:

\begin{equation}
A_{\rm Kp} \propto \left(\frac{L}{M} \right)^{s} \frac{1}{ T_{\rm eff}^{r-1} c_{\rm K}(T_{\rm eff})} \: ,
\label{equ:amps4}
\end{equation}

\noindent
with $c_{\rm K}$ being the bolometric correction factor as a function of effective temperature, 
given by \citet{ballot}:

\begin{equation}
c_{\rm K}(T_{\rm eff}) = \left(\frac{T_{\rm eff}}{5934\rm K}\right)^{0.8} \: .
\end{equation}

\noindent
The reason for choosing to correct the model amplitudes rather than the observed values is that 
effective temperatures for most of the stars in our sample are rather uncertain, and we hence 
prefer not to perform the correction on the observed amplitudes. 

\begin{figure}
\begin{center}
\resizebox{\hsize}{!}{\includegraphics{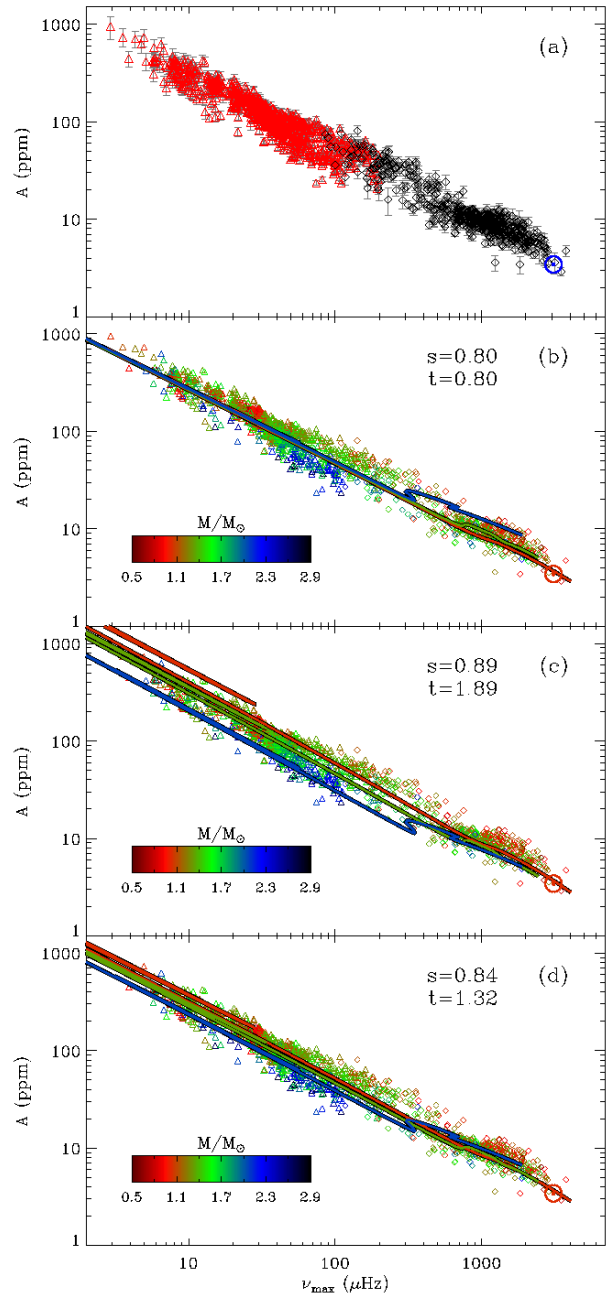}}
\caption{(a) Oscillation amplitude versus \numax\ for the entire \kep\ sample. Symbol types and 
colors are the same as in Figure 1. The position of the Sun is also marked.
(b) Same as panel (a) but with asteroseismic masses color-coded. Error bars have been omitted for 
clarity. Solid lines show the 1.0, 1.3 and 2.0\,$M_{\sun}$ solar-metallicity ($Y=0.273$, $Z=0.0198$)
models scaled using Equation (\ref{equ:amps4}) with typical values of 
$r=2$ and $s=0.8$. (c) Same as panel (b) but using Equation (\ref{equ:amps3}) 
with coefficients determined by fitting the 
model tracks to observations. (d) Same as panel (c) but with best fitting coefficients determined by 
comparing observed amplitudes with calculated amplitudes using asteroseismic masses and radii 
(see text for details).}
\end{center}
\label{fig:amp1}
\end{figure}

Figure \ref{fig:amp1}(a) shows the observed amplitudes 
for the full \kep\ sample as a function of \numax. 
The spread is much larger than the 
typical measurement uncertainties. As first noted by \citet{HBS10} and later confirmed by 
\citet{mosser_mixed}, \citet{mosser_energy} and \citet{stello_cluster}, this spread in the 
amplitude-\numax\ relation for red giants is related to a 
spread in mass. To demonstrate this, Figure \ref{fig:amp1}(b)
shows the same plot but color-coded by stellar masses, as calculated in the previous section. 
We observe that, 
particularly for low-luminosity red giants, the higher-mass stars show lower amplitudes 
than lower-mass stars for a given \numax. For unevolved stars we can tentatively identify the same trend, 
although the separation is less clear.

The observation of a mass dependence for a given \numax\ 
indicates that Equation (\ref{equ:amps4}) should be revised to include an 
additional mass dependence. To demonstrate this, solid lines in Figure \ref{fig:amp1}(b) show 
model tracks with different masses scaled using typical values of $r=2$ and $s=0.8$ in Equation 
(\ref{equ:amps4}).
The $(L/M)^s$ scaling clearly fails to reproduce the observed spread on the red-giant branch, 
but predicts a strong mass dependence for unevolved stars, contrary to what is observed.

An obvious way to account for an additional mass dependence is 
to rearrange Equation (\ref{equ:amps4}) as follows:

\begin{equation}
A_{\rm Kp} \propto \frac{L^{s}}{M^{t} T_{\rm eff}^{r-1} c_{\rm K}(T_{\rm eff})}  \: .
\label{equ:amps3}
\end{equation}

\noindent
Note that such a formulation has been introduced by \citet{KB11} (hereafter KB11) and 
has also been used by 
\citet{stello_cluster} in the \kep\ study of cluster red giants.
To evaluate the coefficients that best reproduce our observations, we again used models with 
masses 1.0$M_{\sun}$, 1.3$M_{\sun}$ and 2.0$M_{\sun}$ and calculated their expected amplitudes according 
to Equation (\ref{equ:amps3}) for a given set of $s$, $t$ and $r$. The agreement of models and
observations was evaluated as follows: For each observed amplitude, 
we interpolated each model track to obtain the model amplitude at the observed \numax\ value. We then 
calculated the minimum squared deviation 
of the observed amplitude to the three model amplitudes, normalized by the measurement uncertainty. 
The parameters $s$, $t$ and $r$ were then optimized using a least-squares fit.
Note that for simplicity we have only used model 
values below the tip of the red-giant branch, since an inclusion of later evolutionary stages 
would require the identification of all red-clump stars in our sample which is beyond the scope of this 
paper. 

First tests quickly showed that the strong correlation between effective temperature and luminosity 
made an independent determination of $s$, $t$ and $r$ impossible. Following \citet{stello_cluster}, 
we have therefore fixed the value to $r=2$.
The best-fitting parameters are $s=0.886\pm0.002$ and $t=1.89\pm0.01$, with uncertainties 
estimated by repeating the fitting 
procedure 1000 times using amplitudes drawn from a random distribution with a scatter 
corresponding to the $1\,\sigma$ measurement uncertainties (scaled so that reduced $\chi^2 = 1$ for 
the original data). The parameter
uncertainties were then estimated by calculating the standard deviation of the resulting distribution 
for each coefficient. 

The scaled model tracks in Figure \ref{fig:amp1}(c) using these coefficients 
reproduce a mass spread in amplitude for red giants, 
with higher-mass stars showing lower 
amplitudes for a given \numax, as observed by \citet{HBS10}, \citet{mosser_mixed}, 
\citet{stello_cluster} and \citet{mosser_energy}. 
For unevolved stars, the model amplitudes now show less dispersion but appear to 
systematically underestimate amplitudes for both subgiant and main-sequence stars.

\begin{figure}
\begin{center}
\resizebox{\hsize}{!}{\includegraphics{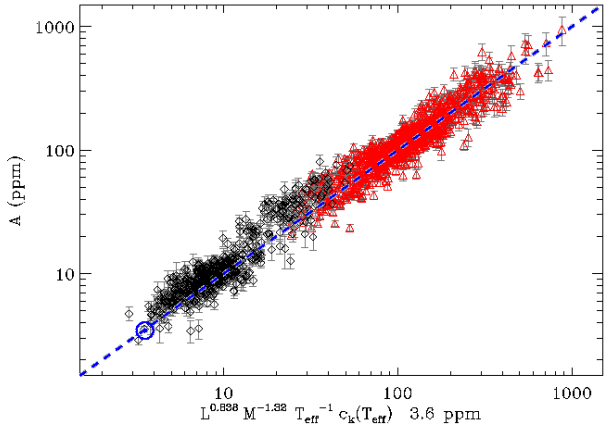}}
\caption{Observed versus calculated amplitudes for the best-fitting coefficients in Equation
(\ref{equ:amps3}), 
using radii and masses determined using the scaling relations for \numax\ and \Dnu. 
The dashed blue line shows the 1:1 relation. The position of the Sun is also shown.}
\label{fig:amp3}
\end{center}
\end{figure}

The method presented above assumes that the excess spread of amplitudes at a given 
\numax\ is entirely due to a spread in stellar mass. 
A more sophisticated approach is to account for the actual mass distribution by directly using masses 
and radii of the sample estimated using 
\numax\ and \Dnu\ in Equations (\ref{equ:nmax}) and (\ref{equ:dnu}). 
We again used the effective temperatures from KIC and 
evaluated Equation (\ref{equ:amps3}) for each value of $s$ and $t$ directly using these 
asteroseismic masses and radii. 
The corresponding best-fitting parameters in this case were $s=0.838\pm0.002$ and $t=1.32\pm0.02$, 
and the scaled model tracks with these parameters are shown in Figure \ref{fig:amp1}(d). 
The spread of the model tracks on the red giant branch is now considerably reduced, better 
matching the observations, 
and the overall fit for less evolved stars is also improved. 

Figure \ref{fig:amp3} compares the observed and calculated amplitudes using $s=0.838$ and 
$t=1.32$, showing good agreement along the 1:1 line over the entire range of \numax.
The residuals between observed and calculated amplitudes over the full range of \numax\ 
show a standard deviation of about 25\%, which is consistent 
with the typical uncertainties in the adopted stellar properties (estimated from propagating the 
uncertainties in \numax\ and \Dnu\ and assuming an uncertainty of 200\,K in $T_{\rm eff}$) 
and the measurement uncertainties in the observed amplitudes. 
We do see an overall 
bias of about 9\%, with calculated amplitudes being systematically underestimated 
when scaling from the solar value. This bias is stronger for unevolved stars with $\numax>300\muHz$ 
(15\%) than for evolved stars $\numax<300\muHz$ (7\%). An 
explanation for this bias might be additional physical differences between the Sun and our 
\kep\ sample, which influence oscillation amplitudes but have not yet been taken into account.

Our best-fitting values for $s$ and $t$ using asteroseismic masses and radii are significantly 
different to \citet{stello_cluster}, 
who found $s=0.90\pm0.02$ and $t=1.7\pm0.1$ using independently determined properties of cluster 
red giants in the \kep\ field. This difference is presumably due to the fact that our sample includes 
a much wider range of evolutionary states. Indeed, repeating our analysis using only red giants in 
a \numax\ range similar to that used by \citet{stello_cluster} yields coefficients which are in 
better agreement. Remaining differences could potentially be due to  
metallicity effects which have been suggested to have a significant influence on 
oscillation amplitudes \citep{samadi} or systematic errors when 
estimating stellar properties through the direct method of using \numax\ and \Dnu. Nevertheless, 
all methods confirm that scaling relations with a separate mass and luminosity 
dependence better reproduce the observed amplitudes from the main sequence to red giants.

\subsection{The KB11 Scaling Relation}

\citet{KB11} recently argued that amplitudes of solar-like oscillations should scale
in proportion
to fluctuations due to granulation. They proposed a revised scaling relation for 
velocity amplitudes:

\begin{equation}
A_{\rm vel} \propto \frac{L \tau_{\rm osc}^{0.5}}{M^{1.5} T_{\rm eff}^{2.25}} \: ,
\end{equation}

\noindent
where $\tau_{\rm osc}$ is the mode lifetime.

Using the same arguments as in the previous section, the KB11
relation predicts photometric amplitudes in the \kep\ bandpass as follows:

\begin{equation}
A_{\rm Kp} \propto \frac{L \tau_{\rm osc}^{0.5}}{M^{1.5} T_{\rm eff}^{1.25+r} c_{\rm k} (T_{\rm eff})} \: .
\label{equ:gran2}
\end{equation}

\noindent
They also suggested the following relation for the granulation power in intensity measured at 
\numax:

\begin{equation}
P_{\rm int}(\nu_{\rm max}) \propto \frac{L^{2}}{M^3 T_{\rm eff}^{5.5}} \: .
\label{equ:gran}
\end{equation}

\noindent
First results on granulation properties of \kep\ red-giant stars 
by \citet{mathur_gran} have shown promising agreement. More recently, 
\citet{mosser_energy} confirmed the proportionality between mode amplitudes and the granulation 
power at \numax\ for a large sample of red giants, but found that the absolute values calculated using 
Equations (\ref{equ:gran2}) and (\ref{equ:gran}) were offset from the observed values.
Here we extend this comparison 
to include also main-sequence stars. 

\begin{figure}
\begin{center}
\resizebox{\hsize}{!}{\includegraphics{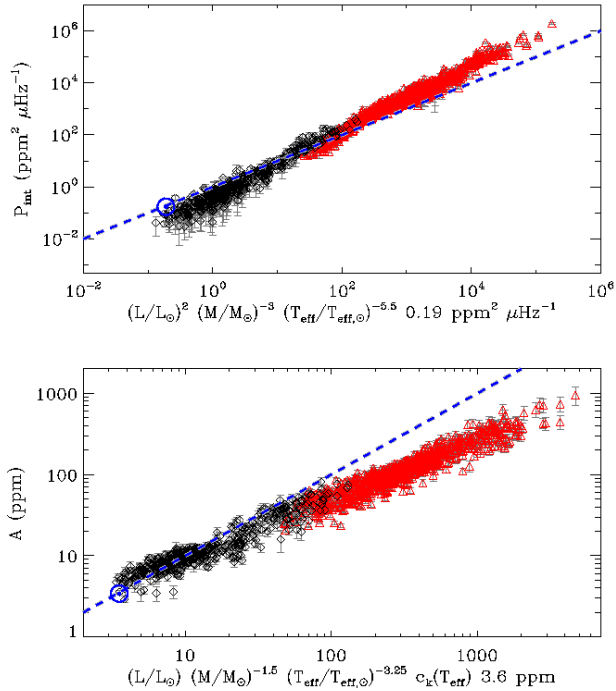}}
\caption{Upper panel: Observed versus calculated granulation power at \numax\ using 
Equation (\ref{equ:gran}). Lower panel:
Observed versus calculated amplitudes using
Equation (\ref{equ:gran2}) and assuming a solar mode lifetime for all stars. 
The dashed blue line shows the 1:1 relation in both panels, and the position of the Sun is marked.}
\label{fig:gran}
\end{center}
\end{figure}

To test the relation, we used the solar granulation power at 
\numax\ determined by our analysis of VIRGO data, which yielded  
a mean value of $P_{\rm int, \sun} = 0.19 \pm 0.02$\,ppm$^{2}\mu$Hz$^{-1}$.
The upper panel of Figure \ref{fig:gran} compares the proposed relation in 
Equation (\ref{equ:gran}), again using asteroseismic masses and radii, as well as effective 
temperatures from KIC. We observe that the calculated values are systematically too high 
for main-sequence stars, and systematically too low for red giants, the latter being in 
agreement with the results by \citet{mosser_energy}. 

In order to compare observed with calculated amplitudes using Equation (\ref{equ:gran2}), 
estimates of the mode lifetimes are required. While it is well established 
that mode lifetimes in giants are significantly longer than in main-sequence stars 
\citep{deridder09,chaplin_lifetimes,hekker10b,HBS10,baudin}, $\tau_{\rm osc}$ is not very well constrained 
for stars spanning such a large range in evolution as considered in our sample. We 
have therefore decided to neglect the influence of $\tau_{\rm osc}$ in Equation (\ref{equ:gran2}) in our study.
The lower panel of Figure \ref{fig:gran} shows the comparison between observed and calculated 
amplitudes using Equation (\ref{equ:gran2}) and assuming solar mode lifetime for all stars. 
We observe that the revised scaling relation works well for 
main-sequence and subgiant stars, but predicts amplitudes that are systematically too high for 
red giants, again in agreement with \citet{mosser_energy}.

We note that the upper panel of Figure \ref{fig:gran} displays an apparent 
systematic difference in the granulation power at \numax\ in the overlapping region of the long-cadence and 
short-cadence sample. This effect is due to a systematic difference of the modelled background noise level 
for stars oscillating close to the long-cadence Nyquist frequency (283\muHz). We have confirmed 
that this has a negligible effect on the measured amplitudes, as is evident by the consistent overlap 
of datapoints in the bottom panel of Figure \ref{fig:gran}.

As noted by \citet{KB11}, the observed differences are not unexpected since the relation does 
not take into account potential differences in the intensity contrast between dark and bright 
regions on the stellar surface. 
While our study confirms the relation between mode amplitudes and granulation power 
for stars ranging from the main-sequence to the red-giant branch, 
it demonstrates that the link between
velocity and intensity measurements is not yet fully understood for amplitude scaling relations.

\subsection{Stellar Activity \& Amplitudes}

The offset between observed and calculated amplitudes for main-sequence stars noted in 
Section 4.1
suggests that there might be an additional physical dependency which is not yet taken into account. 
Increased stellar activity and magnetic fields are known to 
decrease pulsation amplitudes in the Sun \citep{chaplin2000,komm}, and have been suggested as a 
possible cause for the unexpected low amplitudes of oscillations in active stars \citep{hd175726,dall}. 
The first direct evidence for an influence of stellar activity on oscillation amplitudes for a star 
other than the Sun has been found in the F-star HD\,49933 \citep{garcia_science}. 
As discussed by KB11, these amplitude changes may be due to changes in the mode lifetimes.

Recently, \citet{chaplin_apj} reported the discovery for main-sequence and subgiant stars 
observed with 
\kep\ that oscillations are less likely to be detected in stars with higher activity.
This led them to conclude that, just like for the Sun, increased 
stellar activity suppresses mode amplitudes. \kep\ results have furthermore shown that the 
detection rate seems to be considerably lower for subgiant stars 
with $\numax\sim400-900\,\muHz$ \citep{chaplin_science,chaplin_apj2,silva}. This was tentatively 
explained by increased magnetic activity in this stage of stellar evolution, as predicted by 
\citet{gilliland85}. While it is known that some detection bias due to 
instrumental artefacts might be present in the frequency range from 200-500\,\muHz\ \citep{garcia}, 
both discoveries strongly suggest that activity must be considered when calculating amplitudes.

While \citet{chaplin_apj} used a simple detection criterion, we attempt here to confirm and quantify 
their discovery by correlating measured pulsation amplitudes with stellar activity.
Several studies have been devoted to stellar activity and variability in the \kep\ field 
\citep{basri,debosscher,ciardi} with various definitions of activity measures. Since our study 
covers a large range of oscillation timescales and amplitudes, we need to define an activity measure 
which separates variability due to oscillations from variations due to activity, and at the same time 
remains sensitive to local minima and maxima in the light curve. To do this, 
we first smoothed each light-curve with a quadratic Savitzky-Golay filter \citep{sg} with a 
full-width corresponding to 20 times the determined oscillation period, and then measured the 
absolute maximum deviation of the smoothed curve from its mean. We denote this measure of activity as 
$r_{\rm sg}$. Note that we have repeated the analysis with smoothing widths ranging between 10-50 times 
the oscillation period, but found not significant difference in the results. When evaluating $r_{\rm sg}$, it must be kept in mind that our sample includes light curves with two very different 
timebases ($\sim$\,30\,d for short cadence, and $\sim$500\,d for long cadence). The samples 
therefore probe different activity time scales, and have hence been separated in the 
subsequent analysis.

\begin{figure}
\begin{center}
\resizebox{\hsize}{!}{\includegraphics{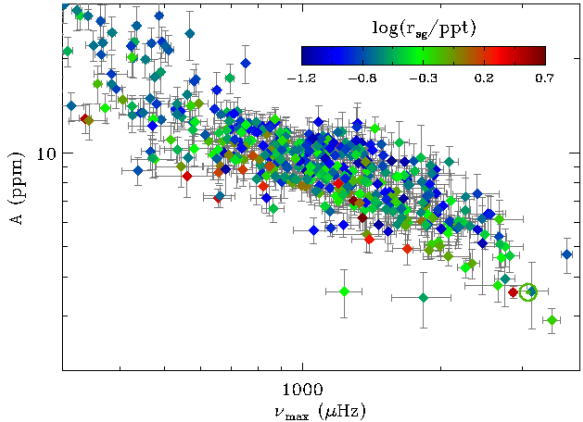}}
\caption{
Amplitude versus \numax\ showing only stars observed in short cadence, with the logarithm of the 
activity range color coded. Red and blue points correspond to high and low activity, respectively. 
The position of the Sun with a color corresponding to the mean solar activity is also shown.}
\label{fig:act1}
\end{center}
\end{figure}

To correct our activity measure for the influence of shot noise, we performed the following 
simulations. For each target light curve, we produced synthetic timeseries including a sinusoidal 
variation with an amplitude corresponding to the measured $r_{\rm sg}$ value. We then added white noise 
with a standard deviation corresponding to the apparent magnitude of the target, which was estimated 
from the minimal noise levels given in \citet{sc} and \citet{lc}. We then measured $r_{\rm sg}$ for the 
synthetic light 
curve as described above, and calculated the difference between this value and the value of $r_{\rm sg}$ 
measured from the 
noise-free time series. This process was repeated 500 times for each star, and the median 
of the resulting distribution was taken as the correction value for the observed $r_{\rm sg}$. 
In summary, it was found that the typical correction values were about 
8\% for short-cadence data, and $<1\%$ for long-cadence data.

\begin{figure}
\begin{center}
\resizebox{\hsize}{!}{\includegraphics{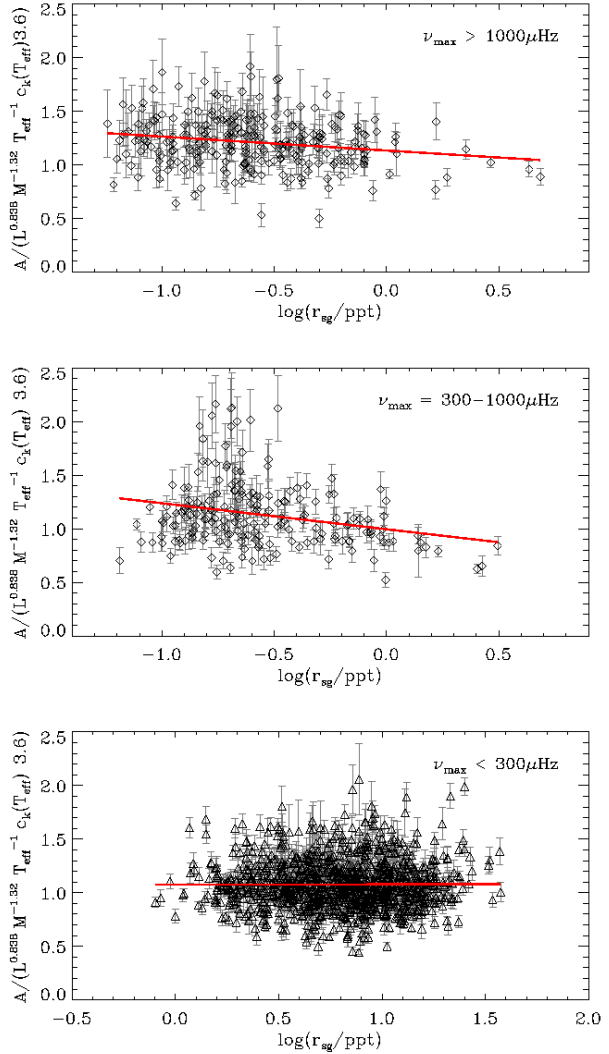}}
\caption{Upper panel: Residuals of observed amplitude divided by the calculated amplitudes using 
Equation \ref{equ:amps3} versus the logarithm of the activity measure 
for all stars with $\numax>1000\muHz$. 
The red solid line shows an unweighted linear fit. Middle panel: Same as the upper panel 
but for all stars with $\numax=300-1000\,\muHz$. Lower panel: Same as upper panel 
but for all long-cadence stars ($\numax<300\muHz$).}
\label{fig:act2}
\end{center}
\end{figure}

Figure \ref{fig:act1} shows the amplitude-\numax\ relation for the short cadence sample, 
with the logarithm of the activity measure color-coded. We see that the most active stars 
(shown in red) have generally lower amplitudes than less active stars (shown in blue).
As found by \citet{chaplin_apj}, 
the \kep\ sample is generally less active than the average Sun 
($r_{\rm sg,\sun}=0.7$\,ppt), 
in agreement with the bias towards lower predicted amplitudes found in the previous section.
To quantify this, we first removed the dependence of oscillation amplitudes on 
stellar properties by dividing them by the calculated amplitudes using 
Equation (\ref{equ:amps3}) with the 
best-fitting coefficients derived in the previous section. The resulting normalized amplitudes 
are plotted versus the logarithm of the activity 
measure in Figure \ref{fig:act2}. 

Figure \ref{fig:act2} is split into three ranges of \numax, roughly dividing stars on the 
main-sequence (top panel), subgiants (middle panel) and red giants (bottom panel). 
The influence of activity on amplitudes appears the strongest for subgiant stars, 
followed by a weak correlation for main-sequence stars and no visible correlation for 
red giants. Unweighted linear fits to the linear-log plots yield slopes of $-0.13\pm0.04$, $-0.24\pm0.07$ 
and $0.00\pm0.02$, respectively. 
While the correlations for unevolved stars are formally significant,
a thorough comparison with other 
methods has shown that the correlation for subgiant stars is only confirmed in three 
out of six methods, while the results for main-sequence and red-giant stars are 
consistent with zero within 3$\sigma$ for all methods. Although this comparison cautions us not to 
draw any definite conclusions, we note that the exponential decrease of amplitude with increasing 
stellar activity for subgiants would 
be in-line with both observational and theoretical evidence for enhanced magnetic activity 
in these stars \citep{gilliland85,chaplin_science}. 
A confirmation of the results presented here will have to await 
longer timeseries (in particular for short cadence data) which will allow a better estimate of the 
activity measure and reduce the uncertainty in the adopted stellar properties through better 
constraints on \numax\ and \Dnu. This will then allow a more in-depth investigation of additional 
physical influences on oscillation amplitudes such as stellar activity and metallicity \citep{samadi}.

\section{The Cool Edge of the Instability Strip}

The cool edge of the instability strip is widely believed to be the dividing line between coherent
pulsations driven by the opacity ($\kappa$) mechanism and solar-like oscillations driven by convection.
Considerable work has been devoted to establishing this boundary both empirically using 
$\delta$\,Scuti and $\gamma$\,Doradus stars \citep[see, e.g.,][]{breger,pamya} as well as 
theoretically \citep[see, e.g.,][]{houdek_inst,houdek2000,xiong,dupret2,dupret05}. The large number of stars for 
which \kep\ has detected oscillations now allows the first test of 
this boundary from ``the cool side'', using solar-like oscillations.

\begin{figure}
\begin{center}
\resizebox{\hsize}{!}{\includegraphics{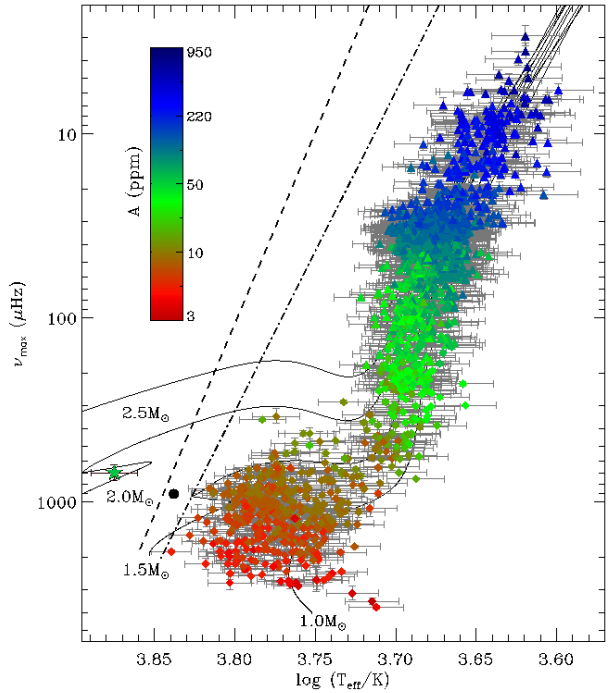}}
\caption{\numax\ versus effective temperature for all 
stars in our sample. Color codes refer to the logarithm of the pulsation amplitude, with blue and 
red colors marking the highest and lowest amplitudes, respectively. The dashed line shows the 
empirically determined cool edge of the instability strip from observations of $\delta$\,Scuti stars. 
The theoretical cool edge for the fundamental 
radial mode of $\delta$\,Scuti pulsations is shown for a 1.7\,$M_{\sun}$ model by \citet{houdek2000} 
(solid circle) and a range of masses by \citet{dupret05} (dashed-dotted line).
Black solid lines are solar-metallicity BaSTI evolutionary tracks with masses as indicated 
in the plot. The star symbol indicates the first detection of hybrid 
solar-like and $\delta$\,Scuti 
oscillations by \citet{antoci}.}
\label{fig:hrd}
\end{center}
\end{figure}

Figure \ref{fig:hrd} shows all stars of our sample in a modified H-R diagram, in which we 
have replaced luminosity with  
$1/\numax$. We also show the empirically determined cool edge of the instability strip 
taken from \citet{pamya} as a dashed line, as well as the theoretical 
cool edge for the fundamental radial mode of $\delta$\,Scuti pulsations for a 1.7$M_{\sun}$ model 
taken from \citet{houdek2000} (filled circle) and for a range of masses by
\citet{dupret05} (dashed-dotted line). 
To convert the lines from the $\log L - \log T_{\rm eff}$ to the $\log \numax - \log T_{\rm eff}$ plane, 
we determined the closest 
matching grid point of evolutionary tracks with different masses to the $\log L - \log T_{\rm eff}$ relation, 
and then fitted a straight line 
to the resulting points in the $\log \numax - \log T_{\rm eff}$ plane. 
This has been done both for BaSTI and a set of ASTEC \citep{astec} 
evolutionary models, and both sets of models yielded consistent results. The empirically determined 
cool edges taken from \citet{pamya} in Figure \ref{fig:hrd} are:

\begin{equation}
\log (T_{\rm eff}/K) =  -0.045 \log (L/L_{\sun}) + 3.893
\label{equ:edge2}
\end{equation}

\begin{equation}
\log (T_{\rm eff}/K) =  0.048 \log (\numax/\muHz) + 3.702 \: .
\label{equ:edge}
\end{equation} 

\noindent
Equation (\ref{equ:edge}) appears
in good agreement with the hottest stars for which solar-like oscillations are observed. 
However, it is important to note that the recent work by \citet{pin} 
and \citet{molenda} suggest that many stars are significantly hotter than the effective temperatures 
given in KIC, with differences of up to 250\,K.
With this in mind, the fact that we observe stars close to the theoretical limit is 
evidence that the separation between the 
classical and stochastic pulsations is probably not a sharp dividing boundary. 
Indeed, we indeed do not observe a gradual 
decrease in amplitude towards the cool edge of the instability strip and
the recent discovery of solar-like and $\delta$\,Scuti oscillations in HD\,187547
by \citet{antoci} using \kep\ data has confirmed the existence of hybrid pulsators (see star symbol in 
Figure \ref{fig:hrd}). Interestingly, 
the observed amplitude for this star is roughly a factor four higher than expected from the 
traditional scaling relations discussed in this paper. This was tentatively explained by the 
increased mode lifetimes which are roughly 4-5 times higher than expected from scaling 
relations based on effective temperature \citep{chaplin_lifetimes,baudin}.

\begin{figure}
\begin{center}
\resizebox{\hsize}{!}{\includegraphics{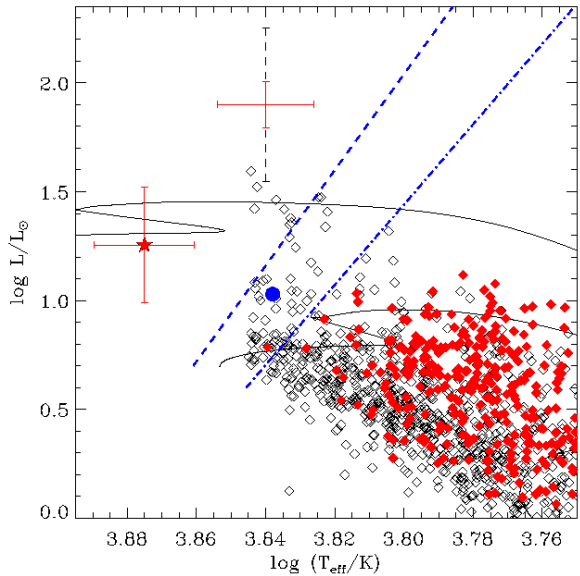}}
\caption{H-R diagram of the low-luminosity region shown in Figure \ref{fig:hrd}, comparing stars 
for which no oscillations have been detected (open black diamonds) with our sample of detections 
(red filled diamonds). 
A typical error bar for both samples is shown in the top region of the plot. 
The dashed line, dashed-dotted line and filled circle are the same as in 
Figure \ref{fig:hrd}. The star symbol indicates the first detection of hybrid solar-like and 
$\delta$\,Scuti oscillations by \citet{antoci}.}
\label{fig:hrd2}
\end{center}
\end{figure}

To test the cool edge of the instability strip more quantitatively, we must account for selection 
effects. Figure \ref{fig:hrd2} shows a close-up of the main-sequence region of the 
HRD, this time plotting luminosity on the ordinate and including stars for which we did not find
evidence for oscillations. Luminosities for stars without detections have been calculated 
based on values in the KIC assuming an uncertainty on the radius of 40\% \citep{verner_kic}, 
while for stars with detections Equations (\ref{equ:nmax}) and (\ref{equ:dnu}) were used. 
Although the error bars are large, we 
observe an excess of hot stars for which no oscillations have been detected, indicating 
that the observed cut-off in detections may not entirely due be due to selection bias. 
We therefore conclude that the empirical red-edge given in Equations (\ref{equ:edge2}) and
(\ref{equ:edge}) gives a good approximation for the transition between opacity driven and 
solar-like oscillations. Further work will be needed to quantify this observation and compare the 
results with theoretical models of excitation and damping of solar-like oscillations, as 
well as possible further detections of hybrid solar-like oscillations and classical pulsation 
in hot stars.

\section{Conclusions}

We have studied global oscillation properties in $\sim$1700 stars observed by \kep\ to test 
asteroseismic scaling relations. Our main findings can be summarized as follows:

\begin{itemize}

\item By comparing evolutionary models with observations, we have 
shown that the scaling relations 
for \numax\ and \Dnu\ are in qualitative agreement with observations for evolutionary stages 
spanning from the 
main-sequence to the He-core burning phase of red giants. The difference in the \Dnu-\numax\ relation 
between evolved and unevolved stars can be explained by different distributions of 
effective temperature and stellar mass, in agreement 
with what is expected from the scaling relations.
A more quantitative test of scaling relations for \numax\ and \Dnu\ will have to await 
the determination of fundamental properties from independent methods such as spectroscopy and 
long-baseline interferometry.

\item We have shown that $(L/M)^s$ scaling for oscillation amplitudes fails to reproduce 
the amplitude-\numax\ relation for red giants.
We have verified that a revised scaling relation using a separate mass and luminosity dependence 
reproduces the observations better, and a relation with $L^{s}$ and $M^{t}$ 
coefficients of $s=0.838\pm0.002$ and $t=1.32\pm0.02$ 
matches amplitudes for field stars 
ranging from the main-sequence to the red clump to 
a precision of 25\% when adopting stellar properties derived from scaling relations.
The calculated amplitudes for main-sequence and subgiant stars in 
our sample are systematically underestimated by up to 15\%, indicating that there might be 
an additional physical dependence which is not yet taken into account.

\item We have investigated the connection of stellar activity with the suppression of 
oscillation amplitudes in main-sequence, subgiant and red-giant stars. We find evidence that the 
effect is strongest for subgiant stars, but caution that these results will have to await 
confirmation with longer time series providing a better estimate of the activity measure and 
reduced uncertainties on stellar properties used to calculated amplitudes. The present data do 
not yield strong evidence that stellar activity contributes significantly to the underestimation of calculated 
amplitudes for main-sequence stars.

\item We have investigated the cool edge of the instability strip using detections of solar-like
oscillations. We find good agreement with the empirically and theoretically determined cool edge using 
$\delta$\,Scuti stars, but note that many stars showing solar-like oscillations may overlap with 
cooler $\delta$\,Scuti stars, in agreement with the recent first discovery of hybrid 
solar-like and $\delta$\,Scuti oscillations by \citet{antoci}. 

\end{itemize}

\acknowledgments
The authors gratefully acknowledge the \kep\ Science Team and everyone involved in the 
\kep\ mission for their tireless efforts which have made this paper possible. Funding for the 
\kep\ Mission is provided by NASA's Science Mission Directorate. 
We thank V. Antoci for helpful comments on the manuscript and discussions on HD\,187547. 
DS and TRB acknowledge support by the Australian Research Council. 
SH acknowledges financial support from the Netherlands Organisation for Scientific Research (NWO). 
HG acknowledges support by the Austrian FWF Project P21205-N16.
JM-\.Z acknowledges the polish Minstry grant N\,N203\,405139.

\bibliographystyle{apj}
\bibliography{references}

\end{document}